\documentclass[%
reprint,
superscriptaddress,
amsmath,amssymb,
aps,
]{revtex4-1}

\usepackage{booktabs}
\usepackage{multirow}

\usepackage{subcaption}
\usepackage{amsmath}

\usepackage{bm} 
\usepackage{graphicx}
\usepackage{dcolumn}
\usepackage{bm}
\usepackage{hyperref}
\usepackage{booktabs}

\begin{document}
	
	\preprint{APS/123-QED}
	
	\title{Non-minimally coupled Einstein-Yang-Mills black holes: periodic orbits and gravitational wave radiation in extreme mass ratio systems}
	\author{Xiuqin He}
	
	\author{Zhaoyi Xu}

	\author{Meirong Tang}%
	\email{tangmr@gzu.edu.cn(Corresponding author)}
	\affiliation{%
		College of Physics,Guizhou University,Guiyang,550025,China
	}%

	
	\begin{abstract}
		Extreme mass ratio inspirals (EMRIs), as a core target for future space-based gravitational wave detection, offer crucial observational grounds for testing strong-field gravitational theories and classifying black holes through their orbital dynamics and gravitational wave radiation characteristics. This study systematically investigates the periodic orbit characteristics and gravitational wave radiation properties of EMRIs in the spacetime of a non-minimally coupled Einstein-Yang-Mills (EYM) black hole. The results show that as the magnetic charge parameter \(Q\) and the non-minimal coupling constant \(\xi\) increase, the radii of the marginally bound orbit (\(r_{\text{MBO}}\)) and the innermost stable circular orbit (\(r_{\text{ISCO}}\)) decrease significantly, with corresponding reductions in the orbital energy \(E\) and angular momentum \(L\). Furthermore, the allowable parameter space of energy and angular momentum (\(E\)-\(L\)) for bound orbits shifts towards the left. We then plot typical periodic orbits through orbit classification, finding that the non-minimal coupling effect suppresses the orbital contraction induced by the magnetic charge, leading to degeneracy of orbits with different \(Q\) values towards the Schwarzschild case, as well as phase shifts, amplitude enhancement, and period shortening in the gravitational waveforms with increasing \(Q\) and \(\xi\). These results provide theoretical predictions for distinguishing different black hole models through future gravitational wave observations.
		
		\begin{description}
			\item[Keywords] Non-minimally coupled Einstein-Yang-Mills black hole; Extreme mass ratio inspirals (EMRIs); Periodic orbits; Gravitational wave radiation.
		\end{description}	
		
	\end{abstract}

	\maketitle
	
	
	\section{\label{sec:level1}Introduction}
	It is a well-established fact that the predictions of general relativity are consistently verified by astronomical observations, which have been indirectly confirmed by astronomical observations through their gravitational or electromagnetic influences on surrounding matter \cite{Lutfuoglu:2025ljm,You:2025ixj}. The recent development of multi-messenger astronomy—supported by gravitational wave (GW) observations from LIGO/Virgo \cite{Lutfuoglu:2025ljm}, direct imaging of black hole (BH) shadows by the Event Horizon Telescope (EHT), and electromagnetic measurements of accretion disk emissions \cite{Gong:2025mne, Wang:2025lpj} —has furnished new multi-messenger approaches for investigating strong-field gravity, verifying general relativity, and probing the spacetime geometry of BHs \cite{Gong:2025mne,Wang:2025lpj, Kondo:2026lqg}.
	
	Within the framework of numerous gravitational theories, general relativity, as a classical description, typically limits its field equations to second-order forms. However, a series of modified gravity theories that go beyond general relativity, by introducing higher-order or more complex tensor coupling terms, may yield new physical effects distinct from the traditional theory \cite{Lutfuoglu:2025ljm,Lee:2018zym,Zi:2026zpw, Ahmed:2025shr,Zhang:2025wni,Zhao:2024exh}. Among these, non-minimally coupled theories, particularly those incorporating non-minimal coupling of gauge fields (such as the Yang-Mills field), constitute an important class of extended models \cite{Lutfuoglu:2025ljm}. These theories couple gravity to the gauge fields describing fundamental particle interactions in a non-trivial manner, motivated potentially by grand unified theories in high-energy physics or low-energy effective approximations of certain quantum gravity schemes. A key theoretical implication is that such modified theories often predict black hole solutions that deviate from the classical ``no-hair theorem(NBT)". The standard NBT \cite{Wang:2025lpj,Kondo:2026lqg,Lee:2018zym} states that a stationary BH is solely determined by only three parameters: $M$, $L$, and $Q$\cite{Meng:2024cnq}. In contrast, BHs in modified theories may carry additional ``hair" parameters (such as $Q$, $\xi$, etc.), which encode information about the BH's microscopic structure.
	
	In this theoretical context, EMRIs \cite{Ahmed:2025shr,Zhang:2025wni}, as a yet-to-be-directly-detected but highly promising source for space-based observations, have garnered significant attention. Such systems typically comprise a supermassive BH with a star mass compact object, like a stellar BH or neutron star, in orbit around it \cite{Zhao:2024exh}. Given the extreme disparity in mass, the centrally located supermassive BH dictates the orbital dynamics of the system long before merger, allowing us to model the stellar-mass object as a test particle on a geodesic trajectory. More importantly, EMRI systems \cite{Wang:2025lpj,Kondo:2026lqg,Lee:2018zym} can undergo hundreds of thousands of slow inspirals before merger, with their GW signals carrying extremely rich and precise information about the geometry fabric of the central supermassive BH, making them an ideal laboratory for testing strong-field gravity theories and exploring the nature of BHs \cite{Tu:2023xab,Hua:2026kvw,Al-Badawi:2023emj}. The EYM BH solutions under non-minimal coupling, while resembling Schwarzschild geometry at large distances, exhibit significant deviations near the event horizon. By studying the periodic trajectories and GW emission of EMRI systems\cite{Al-Badawi:2023emj,Zhao:2024exh}, this paper aims to identify the differences between these models, providing theoretical support for probing the structure near black hole horizons \cite{Zi:2026zpw,Ahmed:2025shr}.
	
	We focus on how two key parameters--\(\xi\) and \(Q\)--affect the stability of bound orbits, orbital precession behavior \cite{Meng:2024cnq}, the classification of periodic orbital path, and ultimately, the properties of the emitted gravitational waveforms by modifying the BH's metric function and effective potential \cite{Munday:2025fdq,DAgostino:2026wln}. The study of periodic orbits holds fundamental significance in GW astronomy \cite{Li:2025eln,Zahra:2025tdo,Zare:2025aek,Ashoorioon:2025ezk,Li:2025sfe,Oliver:2025irg}. Periodic orbits represent the sequence of quasi-stationary orbits that an EMRI system undergoes during its adiabatic evolution, directly determining the long-term behavior of the gravitational wave phase evolution \cite{Huang:2025czx, Meng:2025kzx}. They provide a theoretical foundation for constructing high-precision waveform templates \cite{Haroon:2025rzx,Wang:2025hla,Shabbir:2025kqh, Vargas:2023gvd,DAgostino:2026wln,Zeng:2026ydj}. On the other hand, the stability, frequency ratio, and geometric configuration of periodic orbits are highly sensitive to the geometrical properties of the central BH' spacetime, making them dynamic probes for detecting BH ``hair" \cite{Bruyere:2026gnt}. In recent years, research based on periodic orbit classification methods and rational frequency ratio parameterization has been successfully applied to orbital dynamics and rapid waveform calculations in various BH backgrounds \cite{DAgostino:2026wln,Zeng:2026ydj}, demonstrating the potential of this method in distinguishing different BH models.
	
	Through the systematic analysis outlined above, we will construct a quantitative mapping between the ``hair" parameters \((\xi, Q)\) \cite{Tu:2023xab,Lin:2022ysj,Deng:2021gkx} and the observable quantities of the EMRI system (such as orbital frequency ratio, precession rate, GW phase evolution). This will clarify the observationally distinguishable features of the effects due to $Q$ and arising from $\xi$. This research will provide crucial physical justification and testable predictions \cite{Liu:2020ijk,AraujoFilho:2026rdc,DeAmicis:2026tus} for using EMRI signals captured by future space-based GW detectors (such as LISA, Taiji, Tianqin \cite{Burko:2020gse,Haiman:2017szj,Shabbir:2026qlh}) to test violations of the NBT, identify the type of the central BH, and constrain parameters of modified gravity theories \cite{Wang:2025fmh,Du:2025rty}.
	
	The structure of this paper is as follows: Sec. \ref{sec:level2} introduces the theoretical framework and metric form of non-minimally coupled EYM BH; Sec. \ref{sec:level2} derives the time-like geodesic equations and effective potential in this spacetime; Sec. \ref{sec:level3} analyzes in detail the existence conditions and classification of bound orbits, focusing on the properties of MBO and ISCO; Chapter Sec. \ref{sec:level4} systematically explores the dynamical classification and geometric characteristics of periodic orbits; Section \ref{sec:5} calculates and analyzes the GW radiation waveforms, revealing the modulation effects of \(\xi\) and \(Q\) on the waveforms; eventually, Sec. \ref{sec:6} reviews the paper and analyzes future research directions. Natural units with \(G=c=1\) are used in the present work unless specified otherwise.

	\section{\label{sec:level2}The BH and time-like geodesics}
	The NBT \cite{Biswas:2026mnm} explicitly states that a stationary BH can be totally expressed by \(M\), \(Q\), and \(L\). However, when non-minimally coupled matter fields exist outside the black hole, its external spacetime structure may carry ``hair" information beyond the aforementioned parameters, thereby violating the classical formulation of the no-hair theorem. The non-minimally coupled EYM theory is a typical example of such a scenario. This theory introduces non-minimal interaction terms in the coupling between Einstein gravity and the YM gauge field, thereby introducing new ``hair" in the BH solution--the additional spacetime structure characterized jointly by \(\xi\) and \(Q\). This theoretical framework offers an important modeling basis for studying hairy BH beyond the NBT and their associated dynamics.
	
	The line element corresponding to the static, spherically symmetric BH solution within this theoretical framework can be formulated as follows\cite{Lutfuoglu:2025ljm}
	\begin{equation}
		ds^2 =r^2\left(d\theta^2 + \sin^2\theta d\phi^2\right) -f(r) dt^2 + \frac{1}{f(r)} dr^2 ,
		\label{eq:1}
	\end{equation}
	where \(f(r)\) is given by
	\begin{equation}
		f(r) = 1 + \left(\frac{r^4}{r^4 + 2\xi Q^2}\right)\left(\frac{Q^2}{r^2} - \frac{2 M}{r}\right).
		\label{eq:2}
	\end{equation}
	Here, \(M\) serves as the fundamental source term of the spacetime gravitational field, dictating the global gravitic strength of a BH; \(Q\) is the magnetic charge parameter of the YM gauge field, acting as the ``charge" parameter of the non-Abelian gauge field, modifying spacetime curvature through the metric function, and degenerating to the Schwarzschild solution when \(Q = 0\); \(\xi\) is the non-minimal coupling coefficient, which modulates interaction strength between spacetime curvature and the YM field. When \(\xi = 0\), it recovers the minimally coupled EYM theory. The larger the \(\xi\), the more significant the metric deformation near the event horizon. The extra characteristic of the non-minimally coupled EYM BH is a finite number of observable parameters (Yang-Mills magnetic charge, non-minimal coupling constant) \cite{Zhao:2024exh}, rather than infinitely many degrees of freedom.
	
	After clarifying the non-trivial modulating effect of the parameters on the BH's spacetime structure and its finite degrees of freedom theoretical framework, we can recognize that such hairy BHs not only possess static geometric properties but their dynamic behaviors are also profoundly influenced by \(\xi\) and \(Q\). Particularly in strong-field, multi-orbit dynamic processes such as EMRIs, this spacetime curvature effect shaped by both \(xi\) and \(Q\) will leave unique ``fingerprints" in the periodic orbital motion of test particles and the gravitational wave signals they radiate. Therefore, to systematically explore the periodic orbit characteristics of EMRIs in the background of non-minimally coupled EYM BHs and their observable GW radiation features, we must first ensure that the studied spacetime is physically self-consistent and free of singularity problems. This requires us to strictly limit the BH's parameter range to the physical BH region where event horizons exist, avoiding the unphysical scenario of naked singularities. Consequently, determining the condition for extremal BHs and defining their parameter boundaries becomes the primary theoretical prerequisite for this study. The condition for an extremal BH is \cite{Tu:2023xab,Hua:2026kvw}:
	\begin{equation}
		\left\{
		\begin{array}{c}
			f(r)=0\\
			\left.\frac{df(r)}{dr}\right|_{r=r_h}=0
		\end{array}
		\right..
		\label{eq:3}
	\end{equation}
	
\begin{figure*}
		\centering
		\includegraphics[width=1\linewidth]{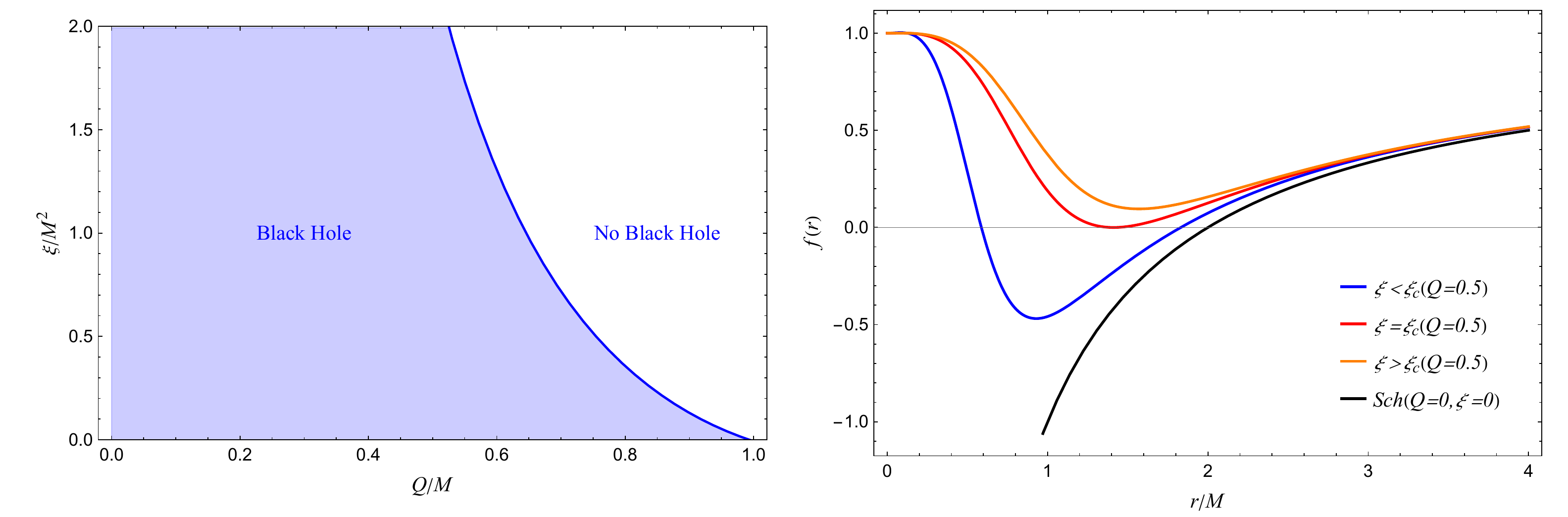}
		\caption{Existence of event horizons in the background of non-minimally coupled EYM BHs (the left panel shows the allowed range of $\xi$ and $Q$ for BH existence, the right panel shows the existence of event horizons).}
		\label{fig:13}
	\end{figure*}

	Numerically solving the above equation system indicates that the extremal BH state is not determined by a single parameter but corresponds to a critical curve \(\xi = \xi_c(Q)\) in the parameter space of \(\xi\) and \(Q\). Here, it is defined that the hairy parameters \((\xi, Q)\) reach a critical value \(\xi_c, Q_c\), i.e., \(\xi = \xi_c\) or \(Q \leq Q_c\). This condition corresponds to the extremal BH case. Fixing \(M = 1\) (in natural units), we select different values of \(\xi\) and \(Q\) to explore the physical effects of this BH. We define the hairy parameters corresponding to the extremal BH as \(\xi = \xi_c\) or \(Q \leq Q_c\), at which point the BH is in a critically coupled state. The Cauchy and event horizon coincide. For $\xi \leq \xi_c$ , $Q \leq Q_c$, the hairy structure remains stable and persistently affects the spacetime surrounding the BH via field coupling \cite{Lutfuoglu:2025ljm}. When \(\xi = 0\), the static ultimate EYM BH degenerates to the Abelian field in the gauge field limit, approximating an RN BH; when the charge is further stripped away (\(Q = 0\)), it degenerates to the Schwarzschild BH. The black line in Fig. \ref{fig:13} shapes the Schwarzschild BH case \cite{Al-Badawi:2023emj}. From the plotted results, it is straightforward to discern that the BH hair causes the event horizon to have a smaller radius than that of the Schwarzschild BH. When \(\xi > \xi_c\), no BH exists at this point, but a naked singularity appears. Different \(Q\) values correspond to different \(\xi_c\) values (corresponding to the solid red trace in the diagram, where \(\xi = \xi_c\)).
	To more intuitively show the allowed parameter region for BH existence, Fig. \ref{fig:13} systematically presents the dependence between \(\xi\) and \(Q\), establishing a clear ``spatial map" for defining the physical research scope of this paper. The shape of the critical curve reveals a profound interaction between the two key parameters. It can be observed that as \(Q\) increases, the two horizons of the sub-extremal BH gradually merge into a single horizon of the extremal BH; simultaneously, increasing \(\xi\) reduces the value of the critical charge \(\xi_c\) (i.e., the left panel of the figure), essentially because the coupling effect counteracts part of the distortion caused by the ``hair". The region below the critical curve \(\left(\xi \leq \xi_c(Q)\right)\) is the physical region where black holes exist. The domain above the threshold \(\left(\xi > \xi_c(Q)\right)\) corresponding to the solid red trace in the diagram — entails the dissipation of the BH horizon and the onset of a naked singularity, a region with no physical validity.
	
	Within the bounds of the above parameter restrictions, we conduct a study on the EMRI system where the stellar-mass compact object is regarded as a test particle—its gravitational perturbations on the supermassive BH spacetime are deemed negligible. By restricting the test particle’s orbital motion to the equatorial plane with \(\theta = \frac{\pi}{2}\), the corresponding Lagrangian for its geodesic motion \cite{Zhao:2024exh,Meng:2024cnq} is given by
	\begin{equation}
		2\mathcal{L} = g_{\mu\nu} \dot{x}^\mu \dot{x}^\nu = -f(r) \dot{t}^2 + \frac{1}{f(r)} \dot{r}^2 + r^2 \dot{\phi}^2.
		\label{eq:4}
	\end{equation}
	Where \(\dot{x}^\mu = \frac{d x^\mu}{d\tau}\) (\(\tau\) is the affine parameter) represents the particle's four-velocity components. Since the Lagrangian above does not explicitly contain time \(t\) or the angular coordinate \(\phi\), according to Noether's theorem, the particle's corresponding $E$ and $L$ are conserved \cite{Zi:2026zpw}
	\begin{align}
		E &= \frac{\partial\mathcal{L}}{\partial\dot{t}} = g_{tt} \dot{t} = f(r) \dot{t}, \label{eq:5} \\
		L &= -\frac{\partial\mathcal{L}}{\partial\dot{\phi}} = -g_{\phi\phi} \dot{\phi} = -r^2 \dot{\phi}. \label{eq:6}
	\end{align}
	
	Within the framework of general relativity, for a massive test particle moving along a timelike geodesic, its four-velocity obeys the normalization condition, namely,
	\begin{equation}
		g_{\mu\nu} \dot{x}^\mu \dot{x}^\nu = -1.
		\label{eq:7}
	\end{equation}
	Substituting the non-minimally coupled EYM BH metric and expressions (\ref{eq:5}) and (\ref{eq:6}) into Eq. (\ref{eq:7}) yields
	\begin{equation}
		\dot{r}^2 = E^2 - V_{\text{eff}}.
		\label{eq:8}
	\end{equation}
	This equation provides an intuitive description of the conservation law governing the radial kinetic energy and effective potential energy of the test particle, where \(V_{\text{eff}}\) denotes the effective potential. Its explicit form is given by
	\begin{equation}
		V_{\text{eff}}=\left[1+ \left(\frac{r^4}{r^4+2\xi Q^2}\right)\left(\frac{Q^2}{r^2}-\frac{2 M}{r}\right)\right]\left(1 + \frac{L^2}{r^2}\right).
		\label{eq:9}
	\end{equation}
	This effective potential formula fully incorporates the modifications to spacetime structure by the non-minimally coupled EYM field, (\(Q\)), and \(L\), serving as a core tool for analyzing the orbital dynamics of test particles. When \(r \rightarrow \infty\), the effective potential asymptotically approaches 1, i.e., \(\lim_{r\rightarrow\infty} V_{\text{eff}} = 1\). At this point, if \(E^2 > 1\), the particle can escape to infinity (\(\left.\dot{r}^2\right|_{r\rightarrow\infty} > 0\)). If the particle is to be bound within a certain region, then \(E^2 < 1\). The orbits where test particles exist under this condition are the bound orbits we aim to discuss.

	\section{\label{sec:level3}Test particle bound orbits in non-minimally coupled EYM BH spacetimes}
	In this section, we focus our investigation on two distinct classes of orbits \cite{Meng:2025kzx,Zhao:2024exh,Meng:2024cnq}: the marginally bound orbit (MBO) and the innermost stable circular orbit (ISCO). For the marginally bound orbit, the test particle has a vanishing velocity at spatial infinity and is only barely gravitationally bound to the BH, a condition that is characterized by the energy relation $E^2 = 1$. Orbits of this kind are highly unstable—even an infinitesimal perturbation will alter the particle’s motion, leading it to either plunge into the BH or escape to infinity. The ISCO, by contrast, corresponds to the smallest radial distance at which a test particle can maintain a stable circular motion. Any test body located within this critical radius is likewise in an unstable state, and will either fall into BHs or escape the gravitational bound following a slight perturbation. In the subsequent analysis, we conduct a detailed study on these bound orbital configurations of test particles.
	
	For the MBO of the non-minimally coupled EYM BH, the following relations must be satisfied:
	\begin{equation}
		V_{\text{eff}} = E^2 = 1,
		\label{eq:10}
	\end{equation}
	\begin{equation}
		\frac{d V_{\text{eff}}}{d r} = 0.
		\label{eq:11}
	\end{equation}
	\begin{figure*}{htbp}
		\centering
		\includegraphics[width=1\linewidth]{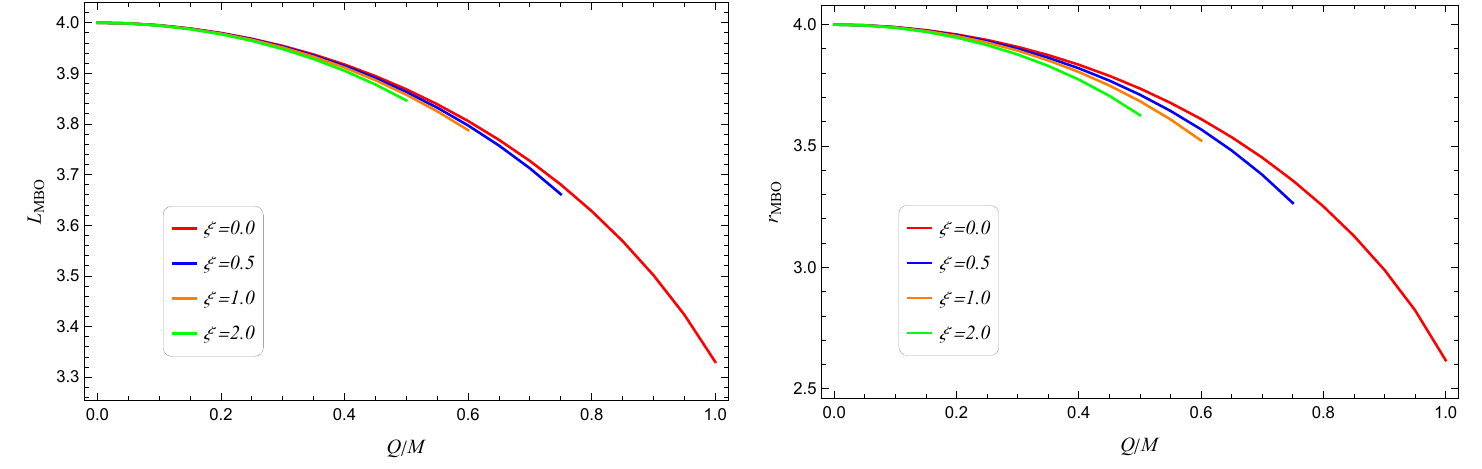}
		\caption{Variation of $r_{\text{MBO}}$ and $L_{\text{MBO}}$ with the parameters.}
		\label{fig:14}
	\end{figure*}
	Numerical simulation is convenient for solving these equations to explore the bound orbits. As shown in Fig. \ref{fig:14}, $r_{\text{MBO}}$ decreases as $Q$ increases. When $Q = 0$, the BH reduces to a Schwarzschild BH, and $r_{\text{MBO}}$ can be calculated as $r = 4M$. As $Q$ increases, the magnetic charge effect causes the radius to decrease, meaning the bound orbit moves closer to the BH. The non-minimally coupling strength parameter $\xi$ amplifies this effect: as $\xi$ increases, the required angular momentum decreases (at $\xi = 0$, corresponding to an RN BH). The non-minimally coupling constant reduces the orbital radius, which aligns with the aforementioned conclusion.
	
\begin{figure*}
		\centering
		\includegraphics[width=1\linewidth]{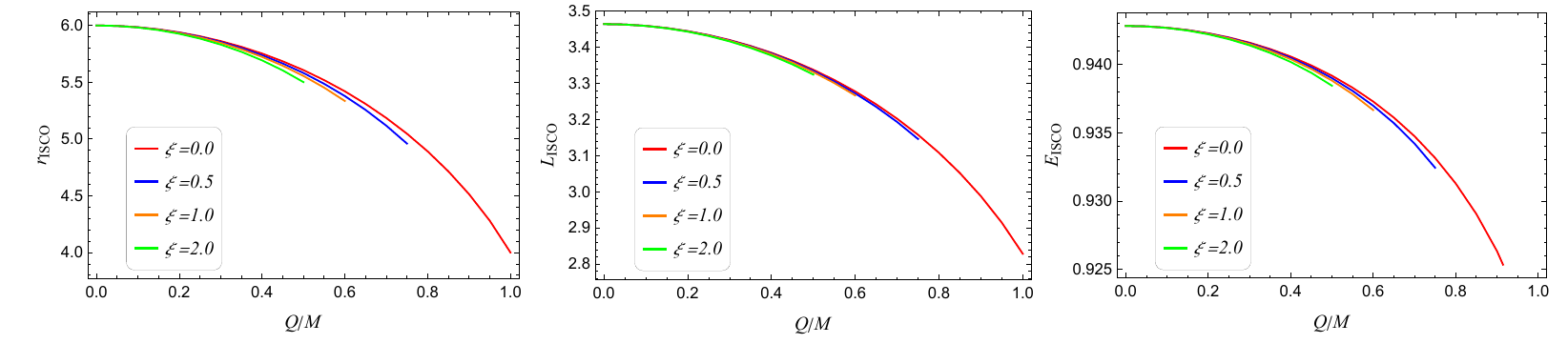}
		\caption{Variation of the ISCO radius $r_{\text{ISCO}}$, $L_{\text{ISCO}}$, and $E_{\text{ISCO}}$ with the parameters.}
		\label{fig:15}
	\end{figure*}

\begin{figure*}
		\centering
		\includegraphics[width=1\linewidth]{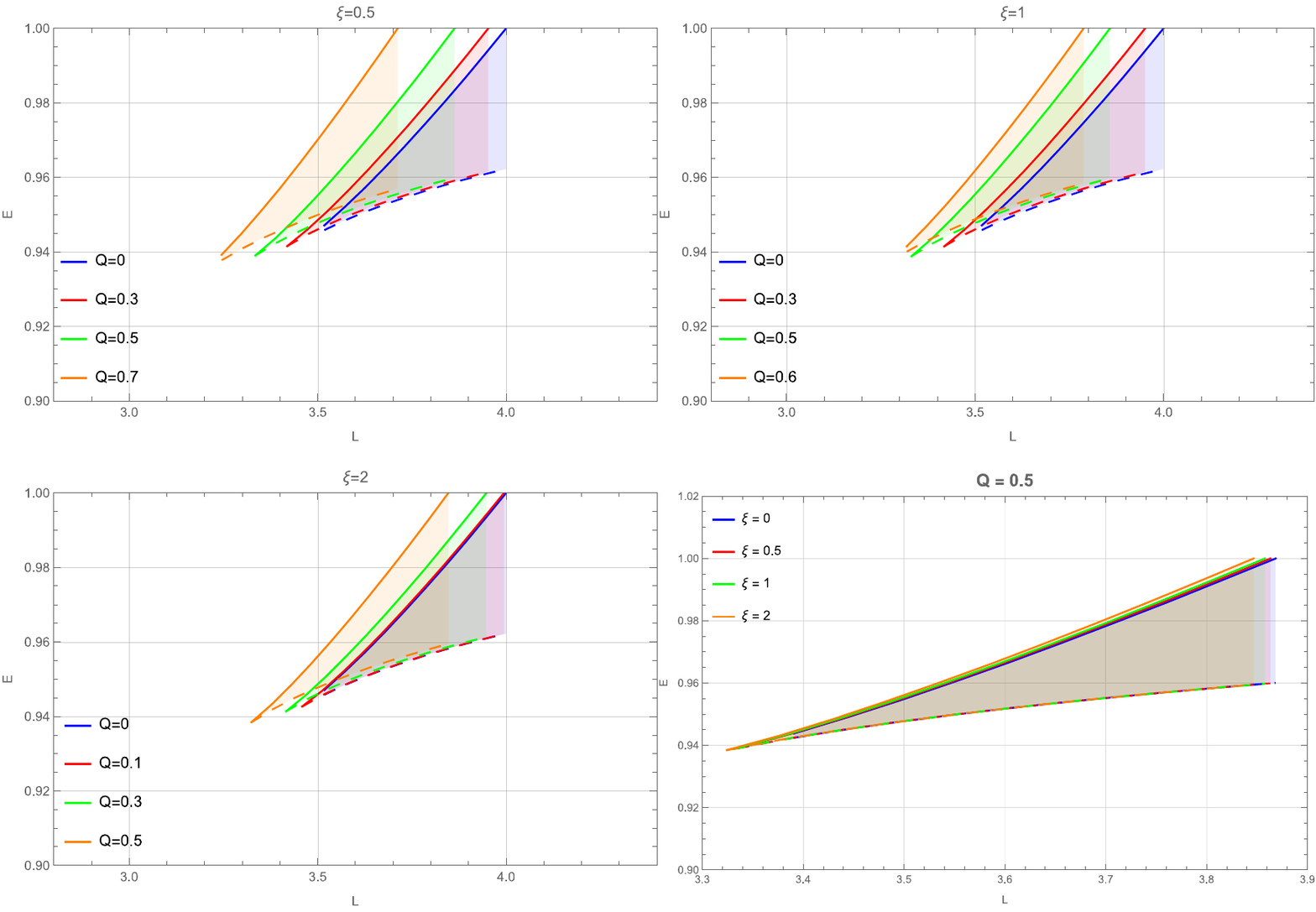}
		\caption{Allowed $E$-$L$ parameter space for particle orbits under different hair parameters $Q$ and $\xi$.}
		\label{fig:61}
	\end{figure*}

\begin{figure}
		\centering
		\includegraphics[width=1\linewidth]{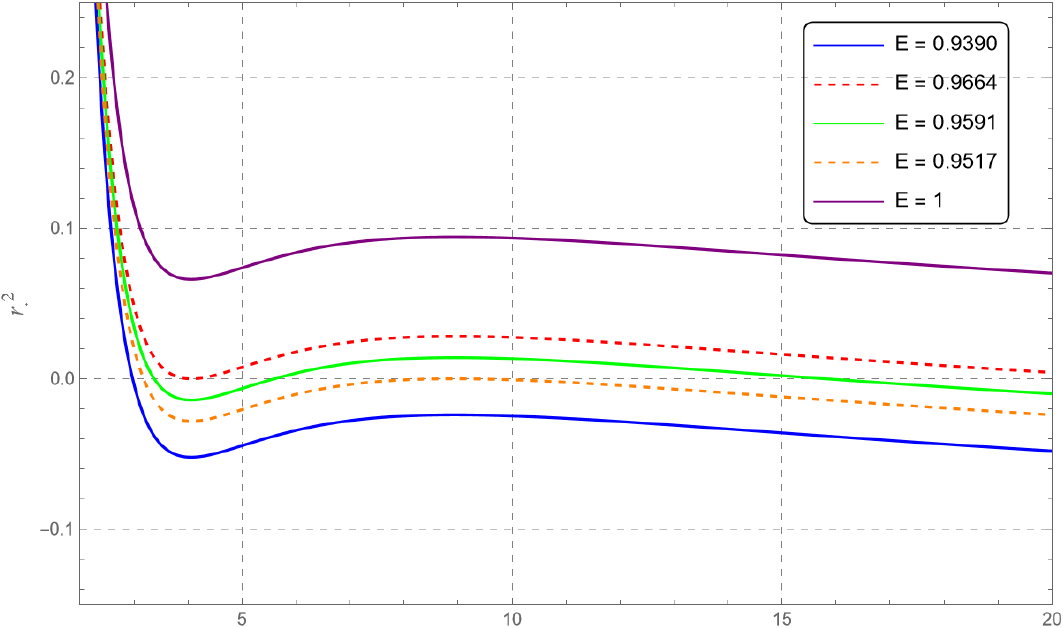}
		\caption{Relationship between the particle's equation of motion $\dot{r}^2$ and the radial coordinate $r$ under different energies.}
		\label{fig:6}
	\end{figure}

	Next, we numerically solve for the ISCO, as shown in Fig. \ref{fig:15}. On this orbit, particles can exist stably. Any additional slight perturbation will cause them to spiral inward inevitably, with the final outcome of plunging into the BH or escaping to the outside. Conditions for the ISCO are
	\begin{equation}
		V_{\text{eff}} = E^2, \quad \frac{d V_{\text{eff}}}{d r} = 0, \quad \frac{d^{2} V_{\text{eff}}}{d r^{2}} = 0.
		\label{eq:12}
	\end{equation}
	By numerically solving for the relationship between the orbital radius $r_{\text{ISCO}}$, $E_{\text{ISCO}}$, and $L_{\text{ISCO}}$ with $Q$, it is found that as $Q$ increases, the $r_{\text{ISCO}}$, $E_{\text{ISCO}}$, and $L_{\text{ISCO}}$ all decrease. When $\xi = 0$, the system reduces to the RN BH parameters, which similarly decrease; when $Q = 0$, it corresponds to the Schwarzschild BH. From Fig. \ref{fig:15}, the ISCO radius is read as $r_{\text{ISCO}} = 6M$. This indicates that the presence of $Q$ and $\xi$ reduces the innermost stable orbit radius, energy, and angular momentum, making the particle closer to the black hole and moving along the orbit near it.
	\begin{figure}
		\centering
		\includegraphics[width=1\linewidth]{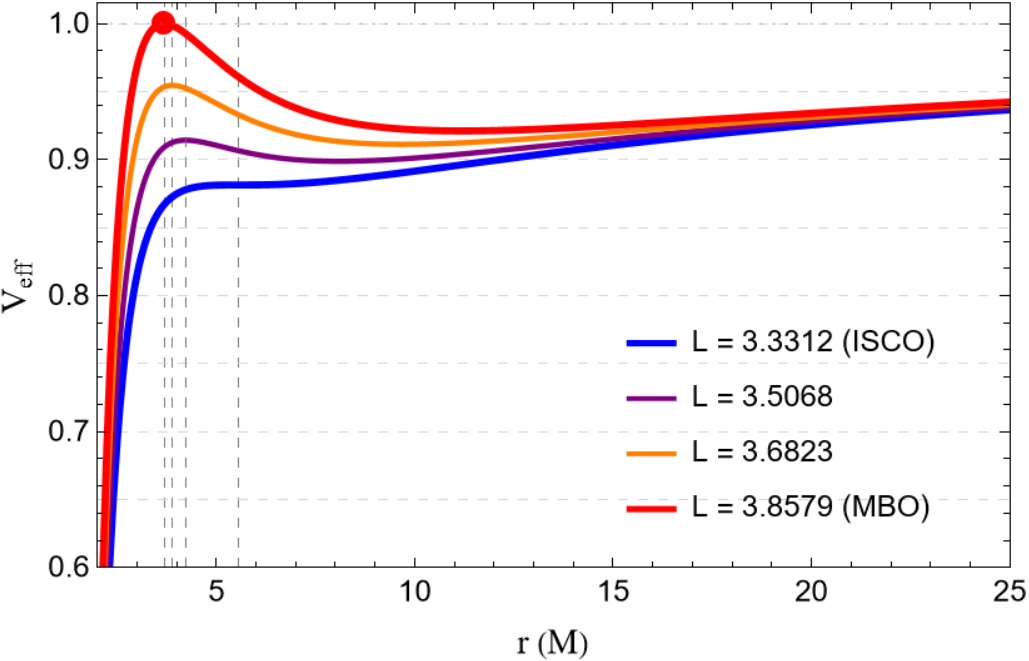}
		\caption{Variation of $V_{\text{eff}}$ with $r$ under different ($Q = 0.5$, $\xi = 1.0$).}
		\label{fig:1}
	\end{figure}
	The motion of a test particle between the ISCO and MBO satisfies Eqs. \ref{eq:10} and \ref{eq:11}. We next examine the dependence of the effective potential on the orbital radius for motion between the ISCO and the MBO. Here, we fix $Q = 0.5$ and $\xi = 1.0$. From Fig. \ref{fig:1}, the red solid line represents the angular momentum of the MBO ($L = 3.8579$, which can be read from Fig. \ref{fig:14}), and the blue solid line represents the angular momentum of the ISCO ($L = 3.3312$, which can also be read from Fig. \ref{fig:13}). The former has two extreme points, while the latter has only one. When the particle moves in the orbit with $3.3312 < L < 3.8579$, its energy lies between $E_{\text{MBO}}$ and $E_{\text{ISCO}}$. The existence of the bound orbit is controlled by $E$ and $L$, and the energy range is related to the choice of angular momentum. For this reason, we conduct an investigation into the dependence of its energy range on angular momentum for different $Q$ values, the findings of which are depicted in Fig. \ref{fig:61}. From left to right, the figures correspond to $\xi = 0.5, 1, 2$ for the relationship between $E$ and $L$, and the last figure corresponds to fixed $Q = 0.3$ with different $\xi$ values. From the results presented in the figure, we find that increasing $L$ leads to an expansion of the energy range within a specific domain. The case where $Q = 0$ corresponds to the Schwarzschild BH. When $\xi$ is large, the image for a certain $Q$ value is similar to that of the Schwarzschild BH, indicating that the coupling effect suppresses $Q$ ; $\xi = 0$ corresponds to the RN BH. From the figure, it can be seen that $\xi$ shifts the image to the left, reducing the maximum and minimum values of $L$, and the larger the $\xi$, the more pronounced the effect. As shown in Fig. \ref{fig:6}, we plot the relationship between $ \dot{r}^2$ and radius $r$ for particle motion under different energies when $Q = 0.4$ and $\xi = 0.5$. Here, $L$ takes the average of the angular momentum of the MBO and the ISCO. From the image, it can be analyzed that when $E > 0.9390$, the particle has one real root, two real roots, three real roots, or no real roots at the zero point of the motion equation. When $E = 0.9517$, $r^2 = 0$ has two real solutions, and between the two real solutions, $r^2 < 0$, indicating that the particle does not exist in a bound orbit. When $E = 0.9390$, there is only one root, and the particle can only fall into the black hole in a specific region, without a bound orbit. When the energy is in the range $0.9517 < E < 0.9664$, there are three roots, and $r^2 > 0$, indicating that the particle has a bound orbit. In the range greater than $0.9664$, the particle can only fall into the black hole from a certain point. The above conclusions are consistent with the conclusions in Fig. \ref{fig:61}.
	
	\section{\label{sec:level4}Periodic Orbital Dynamics}
\begin{figure*}
		\centering
		\includegraphics[width=1\linewidth]{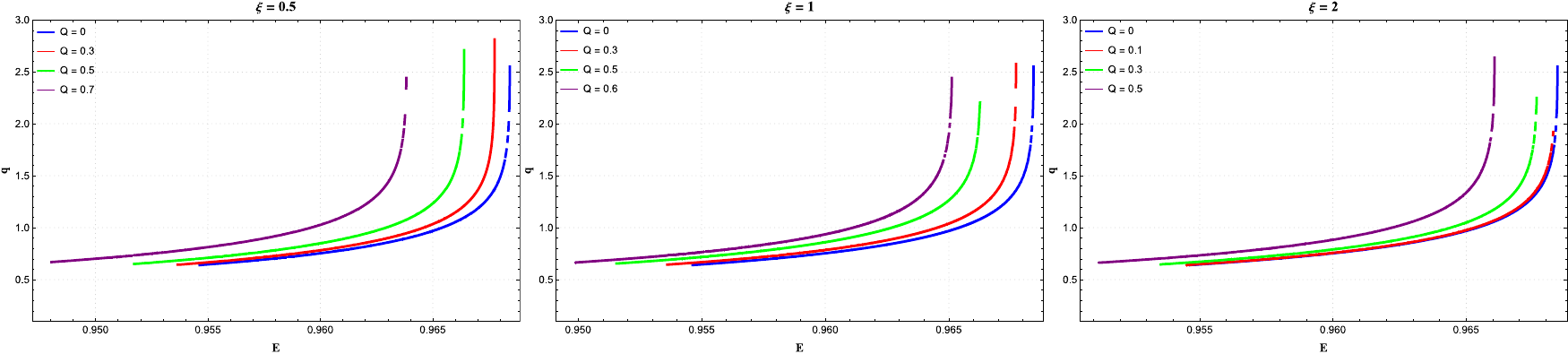}
		\caption{Relationship between $q$ and $E$ for $\varepsilon=0.5$.}
		\label{fig:62}
	\end{figure*}

\begin{figure*}
		\centering
		\includegraphics[width=1\linewidth]{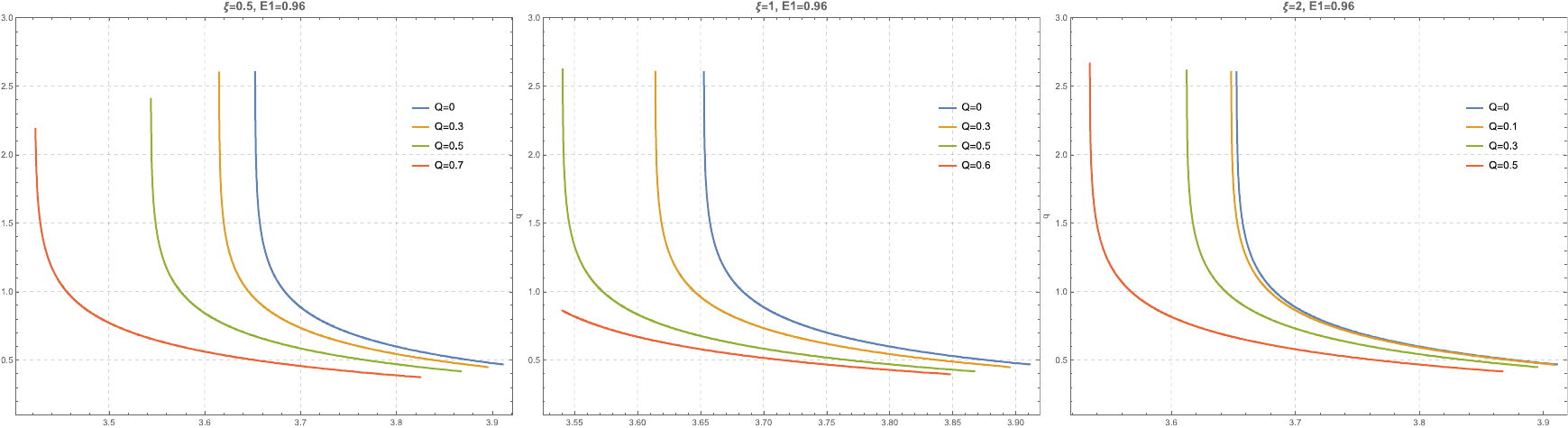}
		\caption{Trend between $q$ and $L$.}
		\label{fig:60}
	\end{figure*}

	In the background of non-minimally coupled EYM black holes, periodic orbits in extreme mass ratio systems exhibit a series of profound and unique physical characteristics. In contrast to systems with standard mass ratios, the configuration composed of a a central supermassive BH (SMBH) and an orbiting stellar-mass object (SMO)—which can represent a stellar companion or a star-like BH (sBH)—serves as an efficient compact source, obeying the mass ratio condition $\frac{M_{\text{SMO}}}{M_{\text{SMBH}}} \leq 10^{-4}$ \cite{Zeng:2026ydj}. The orbital dynamics of this system exhibit new complexities under the influence of $\xi$. In particular, the metric corrections introduced by the non-minimally coupled EYM theory significantly alter the geometric structure around the BH, thereby causing the orbital stability, radial and angular orbital motion, and even the classification and evolution of periodic orbits to exhibit different behavioral patterns compared to classical black hole models (such as Schwarzschild black hole). Therefore, we will adopt the scheme established in the literature to investigate the periodic orbital dynamics of EMRI systems in non-minimally coupled EYM BHs: describing it using a parameter $(z, w, \nu)$ composed of three integers
	\begin{equation}
		q=\frac{\Delta\phi}{2\pi}-1=w+\frac{\nu}{z} .
	\end{equation}
	Where $q$ is called the frequency ratio, $w$ is winding numbers, $\nu$ is vertex numbers, and $z$ is the leaf number. For an irrational value of $q$ the orbital motion of the particle displays behavior analogous to Mercury’s perihelion precession, while the particle's orbit is only periodic when $q$ is rational. This research helps to better explore the properties of non-minimally coupled EYM black holes and the gravitational wave radiation mechanism around them. In the above equation, $\Delta \phi$ can be expressed as
	\begin{equation}
		\Delta \phi = \oint d\phi = 2 \int_{r_1}^{r_2} \frac{d\phi}{dr} dr,
		\label{eq14}
	\end{equation}
	where $r_1$ and $r_2$ represent the pericenter and apocenter radii of the periodic orbit, respectively, determined by the two roots of the orbital eq.  \ref{eq:9} when set to zero. Combining the above formula (\ref{eq14}), the frequency ratio can be rewritten as	
	
	\begin{align}
		q&=\frac{\Delta\phi}{2\pi}-1 \notag\\
		&=\frac{1}{\pi}\int_{r_1}^{r_2} \frac{L}{r^2\sqrt{E^2-\left(1- \frac{2M}{r}+\frac{Q^2}{r^2}\right)\left(1 + \frac{L^2}{r^2}\right)}}dr-1.
	\end{align}
	To better analyze the change of $q$ with respect to $E$ or $L$, we similarly plot the relationship between $q$ and E through numerical simulation in Fig. \ref{fig:62}, where the angular momentum is
	\begin{equation}
		L=L_{\text{ISCO}}+\varepsilon(L_{\text{MBO}}-L_{\text{ISCO}}).
	\end{equation}
	In the figure, we fix $\varepsilon = 0.5$. Analysis of the image shows that: when the energy is within a certain range ($E_{\text{ISCO}} < E < E_{\text{MBO}}$), $q$ slowly increases; when $E$ reaches its maximum value, the $q$ value increases sharply until it diverges. Inclusion of $Q$ leads to an earlier divergence of the curve, and this effect becomes increasingly significant with larger $Q$. In addition, a larger $\xi$ results in a more rapid divergence of $q$.
	In Fig. \ref{fig:60}, the central energy is fixed at $E = 0.96$ (guaranteed value). From the numerical results, we observe that as $L$ decreases, $q$ rises gradually. Upon reaching a critical value of $L$, $q$ exhibits a rapid increase and eventually diverges. A larger $Q$ also leads to a faster divergence of $q$. In Fig. \ref{fig:62} and Fig. \ref{fig:60}, the case $Q = 0$ reduces to the Schwarzschild BH. Analysis of the two images allows us to understand that the non-minimally coupled EYM BH differs from the classical Schwarzschild BH in that $E$ and $L$ of its periodic orbits are smaller.
	
	To systematically study the dynamics of periodic orbits $(z, w, \nu)$ and their gravitational wave radiation characteristics, we select $E$ (Table \ref{tab:energy_periodic_orbits}) and $L$ (Table \ref{tab:periodic_orbit_angular_momenta}) for smaller $q$ values. The reason for investigating periodic orbits with smaller $q$ values is that when $q$ is larger, the lifetime of the orbit is shorter, and it is extremely sensitive to perturbations, which is unfavorable for research \cite{Lin:2022ysj}.
	\begin{table*}[htbp]
		\centering
		\caption{$E$values corresponding to periodic orbits $(z,w,v)$ under different parameters, with fixed $L = \frac{L_{\text{ISCO}} + L_{\text{MBO}}}{2}$.}
		\label{tab:energy_periodic_orbits}
		\begin{tabular}{w{l}{1.5cm}w{l}{2cm}w{l}{2cm}w{l}{2cm}w{l}{2cm}w{l}{2cm}w{l}{2cm}w{l}{2cm}}
			\hline\hline
			$\xi$ & $Q$ & $E_{(1,1,0)}$ & $E_{(1,2,0)}$ & $E_{(2,1,1)}$ & $E_{(3,1,1)}$ & $E_{(3,1,2)}$ \\ \hline
			\multirow{4}{*}{0.5} 
			& 0 & 0.96542534 & 0.96838276 & 0.96802648 & 0.96763354 & 0.96822765 \\
			& 0.3 & 0.96460609 & 0.96769882 & 0.96732157 & 0.96690863 & 0.96753398 \\
			& 0.5 & 0.96292708 & 0.96632873 & 0.96590197 & 0.96544281 & 0.96614073 \\
			& 0.7 & 0.95958917 & 0.96374671 & 0.96318936 & 0.96261390 & 0.96349641 \\
			\hline\multirow{4}{*}{1} 
			& 0 & 0.96542534 & 0.96838276 & 0.96802649 & 0.96763353 & 0.96822765 \\ \
			& 0.3 & 0.96454797 & 0.96766932 & 0.96728695 & 0.96686950 & 0.96750204 \\
			& 0.5 & 0.96270724 & 0.96622052 & 0.96577288 & 0.96529585 & 0.96602243 \\
			& 0.6 & 0.96114426 & 0.96504582 & 0.96452948 & 0.96399208 & 0.96481480 \\
			\hline	\multirow{4}{*}{2} 
			& 0 & 0.96542534 & 0.96838276 & 0.96802649 & 0.96763353 & 0.96822765 \\
			& 0.1 & 0.96532109 & 0.96830072 & 0.96794085 & 0.96754455 & 0.96814393 \\
			& 0.3 & 0.96442907 & 0.96757137 & 0.96721636 & 0.96678959 & 0.96743698 \\
			& 0.5 & 0.96222858 & 0.96599012 & 0.96549438 & 0.96497711 & 0.96576856 \\ \hline\hline
		\end{tabular}
	\end{table*}
	\begin{table*}[htbp]
		\centering
		\caption{$L$ values corresponding to periodic orbits $(z,w,v)$ under different parameters $\xi$ and $Q$.}
		\label{tab:periodic_orbit_angular_momenta}
		\begin{tabular}{w{l}{1.5cm}w{l}{2cm}w{l}{2cm}w{l}{2cm}w{l}{2cm}w{l}{2cm}w{l}{2cm}w{l}{2cm}}
			\hline\hline
			$\xi$ & $Q$ & $L_{(1,1,0)}$ & $L_{(1,2,0)}$ & $L_{(2,1,1)}$ & $L_{(3,1,1)}$ & $L_{(3,1,2)}$ \\ \hline
			\multirow{4}{*}{0.5} 
			& 0 & 3.6835877 & 3.6534056 & 3.6575957 & 3.6618408 & 3.6553017 \\
			& 0.3 & 3.6458956 & 3.6157667 & 3.6199399 & 3.6241755 & 3.6176535 \\
			& 0.5 & 3.5748639 & 3.5444921 & 3.5487081 & 3.5529832 & 3.5463988 \\
			& 0.7 & 3.4559244 & 3.4238300 & 3.4284108 & 3.4329754 & 3.4259180 \\
			\hline\multirow{4}{*}{1} 
			& 0 & 3.6835877 & 3.6534056 & 3.6575957 & 3.6618408 & 3.6553017 \\
			& 0.3 & 3.6452117 & 3.6149135 & 3.6191230 & 3.6233870 & 3.6168185 \\
			& 0.5 & 3.5724757 & 3.5414671 & 3.5458226 & 3.5502055 & 3.5434438 \\
			& 0.6 & 3.5405724 & 3.5405724 & 3.5405724 & 3.5405724 & 3.5405724 \\
			\hline	\multirow{4}{*}{2} 
			& 0 & 3.6835877 & 3.6534056 & 3.6575957 & 3.6618408 & 3.6553017 \\
			& 0.1 & 3.6792667 & 3.6490465 & 3.6532440 & 3.6574955 & 3.6509462 \\
			& 0.3 & 3.6438327 & 3.6131868 & 3.6174713 & 3.6217938 & 3.6151292 \\
			& 0.5 & 3.5675569 & 3.5351455 & 3.5398175 & 3.5444402 & 3.5372824 \\
			\hline	\multirow{2}{*}{0} 
			& 0.3 & 3.6465756 & 3.6166132 & 3.6207510 & 3.6249586 & 3.6184823 \\
			& 0.5 & 3.5772072 & 3.5474348 & 3.5515219 & 3.5556959 & 3.5492771 \\ \hline\hline
		\end{tabular}
	\end{table*}
	The data in Table \ref{tab:energy_periodic_orbits} indicate that for the same $q$ value, when $Q$ increases and $L$ is fixed as $L = \frac{L_{\text{ISCO}} + L_{\text{MBO}}}{2}$, $E$ decreases; Table \ref{tab:periodic_orbit_angular_momenta} shows that for the same $q$ value, when $Q$ increases and $E = 0.96$, $L$ increases with $Q$. Based on this, we plot the periodic orbit trajectories $(z, w, \nu)$ of neutral test particles in the background of the non-minimally coupled EYM BH. The coordinates are defined as $(x, y) = (r\cos\phi, r\sin\phi)$.
	\begin{figure*}
		\centering
		\includegraphics[width=1\linewidth]{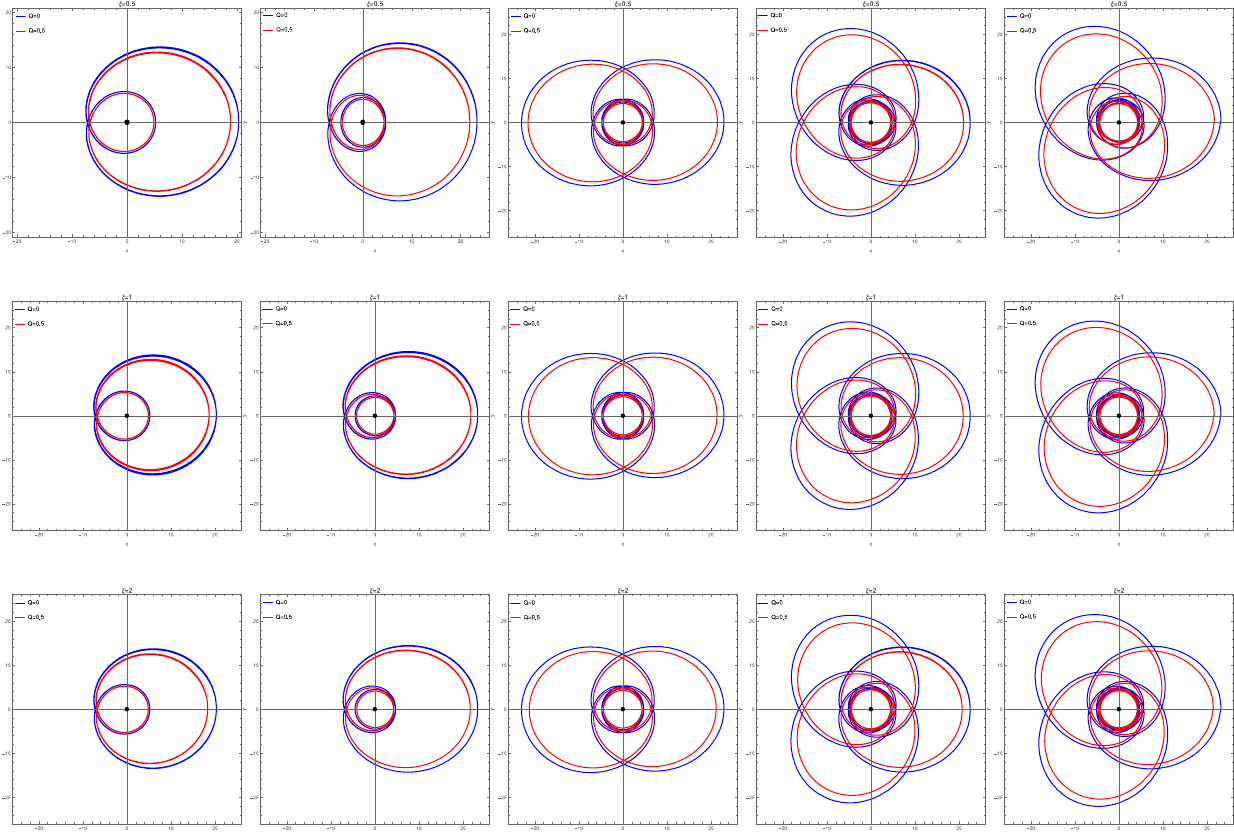}
		\caption{Different periodic orbits for fixed $\varepsilon=0.5$.}
		\label{fig:64}
	\end{figure*}
	\begin{figure*}
		\centering
		\includegraphics[width=1\linewidth]{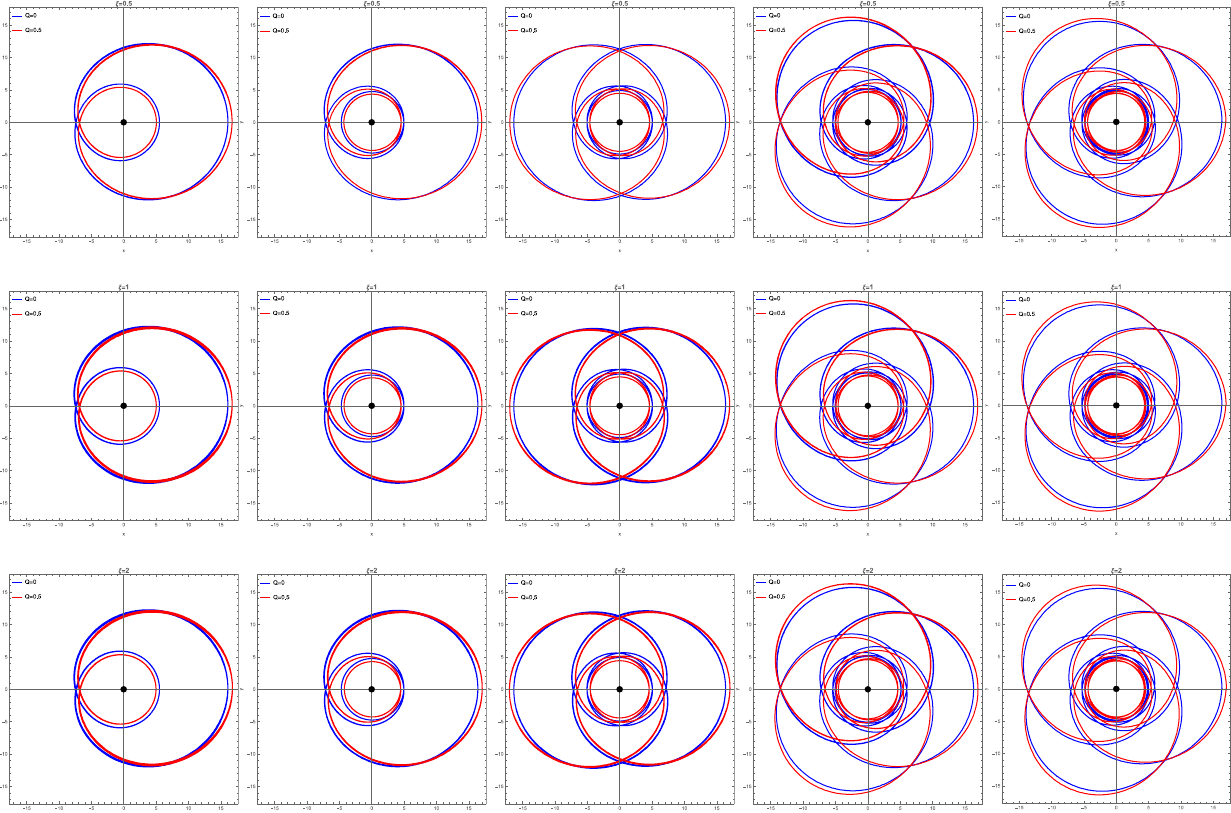}
		\caption{Different periodic orbits with fixed energy $E=0.96$.}
		\label{fig:65}
	\end{figure*}
	\begin{figure*}
		\centering
		\includegraphics[width=1\linewidth]{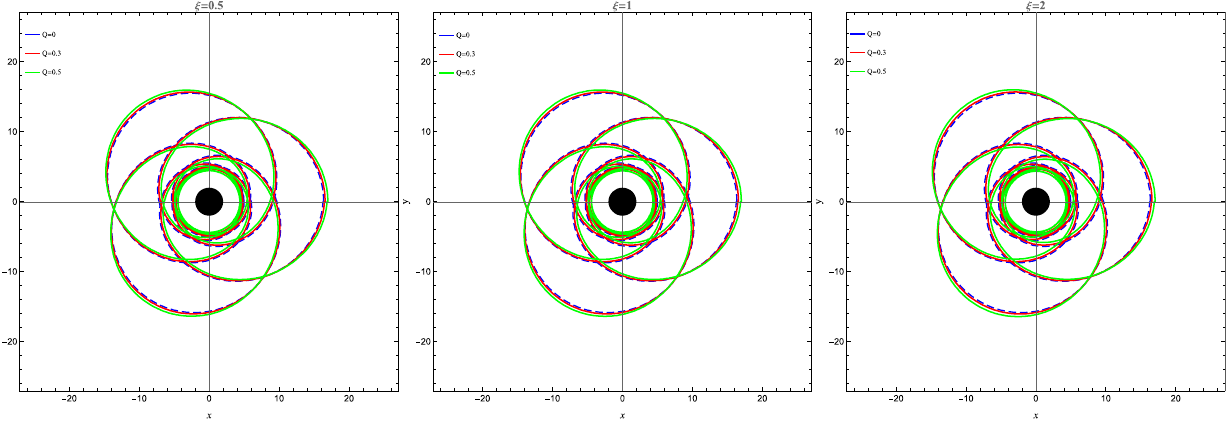}
		\caption{Trajectories of the periodic orbit $(3,1,2)$ under different $\xi$ values.}
		\label{fig:66}
	\end{figure*}
	Fig. \ref{fig:64} and Fig. \ref{fig:65} represent the periodic orbits under different conditions, from top to bottom corresponding to $\xi = 0.5$, $\xi = 1$, and $\xi = 2$. Fig. \ref{fig:64} is drawn from left to right in the order of Table \ref{tab:energy_periodic_orbits}. At this time, $E$ is fixed at $\varepsilon = 0.5$, and $L$ values are taken from Table \ref{tab:energy_periodic_orbits}; in Fig. \ref{fig:65}, $E$ is still fixed at $0.96$, and $L$ values are taken from Table \ref{tab:periodic_orbit_angular_momenta}. The trajectory of the periodic orbit $(3,1,2)$ is plotted numerically in Fig. \ref{fig:65}. Moving from left to right, the images stands for $\xi = 0.5$, $\xi = 1$, and $\xi = 2$. The blue dashed line in the figure represents the Schwarzschild BH. From the information in the figure, it can be seen that: the presence of $Q$ causes the particle trajectory to be larger than that of the Schwarzschild BH when outside the orbit, and this difference becomes more pronounced as $Q$ increases. However, on the inner side of the trajectory, there is a situation closer to the black hole. The increase of $\xi$ causes the particle trajectory to shrink compared to the Schwarzschild BH trajectory, indicating that the presence of $\xi$ can reduce this $Q$ effect.
	The blue curve represents the Schwarzschild BH, and the red curve represents the BH model with $Q = 0.8$. As described in the literature \cite{Zhao:2024exh,Tu:2023xab,Hua:2026kvw}, $z$ represents the number of large circles in the trajectory, $\nu$ represents the number of small circles around the BH, and $w$ represents the number of times the particle passes through the vertex of the orbit. $z$ is proportional to the complexity of the periodic orbit. This study shows that: $\xi$ and $Q$ will change the $(z, w, \nu)$ corresponding to the given $(E, L)$, thereby causing the orbit geometry to deform relative to the Schwarzschild and RN BHs. This deformation is a direct manifestation of the vacuum rate of the non-minimal coupling EYM black hole being modified by the ``hair" parameter. Analyzing the systematic deviation of these periodic orbits provides crucial theoretical predictions and observable distinguishing features for people to test the no-hair theorem in extreme mass ratio inspiral systems, identify the ``hair" type of black holes, and even limit the modification of gravitational theory parameters.
	\begin{table*}[htbp]
		\centering
		\caption{Comparison of orbital parameters between RN black holes and non-minimally coupled EYM black holes.}
		\label{tab:orbital_comparison}
		\begin{tabular}{w{l}{1.5cm}w{l}{1.4cm}w{l}{1.4cm}w{l}{1.4cm}w{l}{1.4cm}w{l}{1.4cm}w{l}{1.4cm}w{l}{1.4cm}w{l}{1.4cm}w{l}{1.4cm}w{l}{1.4cm}}
			\hline\hline
			Parameter & $Q=0.3$ & $Q=0.3$ & $Q=0.3$ & $Q=0.3$ & $Q=0.3$ & $Q=0.5$ & $Q=0.5$ & $Q=0.5$ & $Q=0.5$ & $Q=0.5$ \\
			\hline
			$r_p$ & 4.57651 & 4.6207 & 4.65509 & 4.70943 & 0.13292& 5.66846 & 5.67971 & 5.69076 & 5.71228 & 0.04328  \\
			$r_a$ & 16.5947 & 16.5946 & 16.5946 & 16.5945 & -0.0002 & 16.3474 & 16.3472 & 16.3470 & 16.3466 & -0.0008  \\
			\hline\hline
		\end{tabular}
	\end{table*}
	The Table \ref{tab:orbital_comparison} shows the periodic orbit $(3,1,2)$ trajectories with fixed $Q$ and $E=0.96$. Here, $r_p$ and $r_a$ represent the pericenter and apocenter, respectively. The purpose of introducing these two parameters is to explore the differences in this trajectories under different $\xi$ values. From the table, it can be seen that the presence of non-minimal coupling causes the trajectory to expand outward on the exterior side, but it is closer to this BH on the interior side.
	
	In summary, the ``hair" parameter of the black hole affects the particle's periodic orbit, but compared to the RN and Schwarzschild BH, this spatial curvature effect accumulates at a farther distance, i.e., exhibiting the characteristic of trajectory expansion outward, which is also a good method for distinguishing BHs. The ``hair" of the non-minimally coupled EYM BH produces a unique and observable modulation on the geometric shape of its surrounding periodic orbits. This modulation effect not only provides a new method in theory to test the NBT and identify the ``hair" type of BHs, but also sets a clear theoretical prediction direction for the next generation of gravitational wave astronomy in practice. The analysis of this study shows that through precise comparison of the variation patterns of the pericenter and apocenter of the periodic orbits, as well as the deviation characteristics of the overall orbital geometry relative to Schwarzschild and RN BHs, future high-precision celestial measurements (such as gravitational wave phase analysis) are expected to achieve quantitative measurements of the BH ``hair" parameters ($\xi, Q$), thereby pushing the era of black hole astrophysics from ``static imaging" to ``dynamic diagnosis".
	
	\section{\label{sec:5}Gravitational wave radiation from periodic orbits}
	
	EMRIs are potential important sources for future space-based GW detectors \cite{Zeng:2026ydj}. Within the EMRI configuration, a stellar mass compact body follows periodic orbits about a non-minimally coupled EYM BH, representing a unique GW. As the timescale for orbital evolution is significantly longer than the orbital period, the perturbation exerted by the test particle on the background spacetime can be regarded as a small perturbation. Under these conditions, the adiabatic approximation can be adopted. Within this situation, $E$ and $L$ of the system are approximately conserved over one or several orbital cycles, allowing the orbital motion to be approximated as a series of quasi-geodesic motions, with radiation reaction effects negligible over a single cycle. This treatment not only simplifies calculations but also accurately describes the orbital evolution and GW radiation characteristics of the small mass object in the non-minimally coupled EYM BH background. This method has been well applied in the literature \cite{Munday:2025fdq,Li:2025eln,Zahra:2025tdo,Zare:2025aek}, so we adopt the method verified to be effective in the literature \cite{Ashoorioon:2025ezk,Li:2025sfe,Oliver:2025irg} to compute the corresponding gravitational waveforms.
	
	In the theoretical framework, we employ a numerical kludge scheme, which mainly consists of two key steps: First, we numerically solve the timelike geodesic equations in the background spacetime of the BH, thereby deriving the exact orbital dynamics of the test particle, with a specific focus on the characteristic features of its zoom-whirl motion \cite{Li:2025sfe}; second, we substitute the orbital solution into the quadrupole formula to calculate the corresponding gravitational waveforms. For the GW radiation waveform from a single cycle of orbital motion, we introduce the quadrupole formula \cite{Huang:2025czx,Meng:2025kzx,Haroon:2025rzx}:
	\begin{equation}
		h_{ij} = \frac{4\eta M}{D_L} \left(v_i v_j - \frac{m}{r} n_i n_j\right). 
	\end{equation}
	In the formula, $\eta$ represents the symmetric mass ratio $\frac{M m}{(M+m)^2}$, $D_L$ is the luminosity distance from the extreme mass ratio system to the detector, $M$ is the mass of the non-minimally coupled EYM BH, $n$ denotes the radial unit vector directed from the BH toward the stellar-mass compact object.
	
	The GWs detected by the detector need to undergo appropriate coordinate transformations for easier investigation. Therefore, we introduce a coordinate system centered on the black hole \cite{Wang:2025hla,Shabbir:2025kqh}:
	\begin{align}
		e_X &= [\cos\omega,\; -\sin\omega,\; 0]. \\
		e_Y &= [\cos\iota \sin\omega,\; \cos\iota \cos\omega,\; -\sin\iota]. \\
		e_Z &= [\sin\iota \sin\omega,\; \sin\iota \cos\omega, \;\cos\iota]. 
	\end{align}
	where $\omega$ represents the azimuth angle of the orbital ascending node relative to a reference direction, and $\iota$ is the orbital inclination. In this coordinate system, the two polarization components of the GW can be written in the following form
	\begin{equation}
		h_{+} = -\frac{2\eta M^2}{D_L r} (1 + \cos^2\iota) \cos(2\phi + 2\zeta),
	\end{equation}
	\begin{equation}
		h_{\times} = -\frac{4\eta M^2}{D_L r} \cos\iota \sin(2\phi + 2\zeta).
	\end{equation}
	In the formulas, $\phi$ represents the orbital phase angle, and $\zeta$ represents the orbital dimension angle.
	\begin{figure*}
		\centering
		\includegraphics[width=1\linewidth]{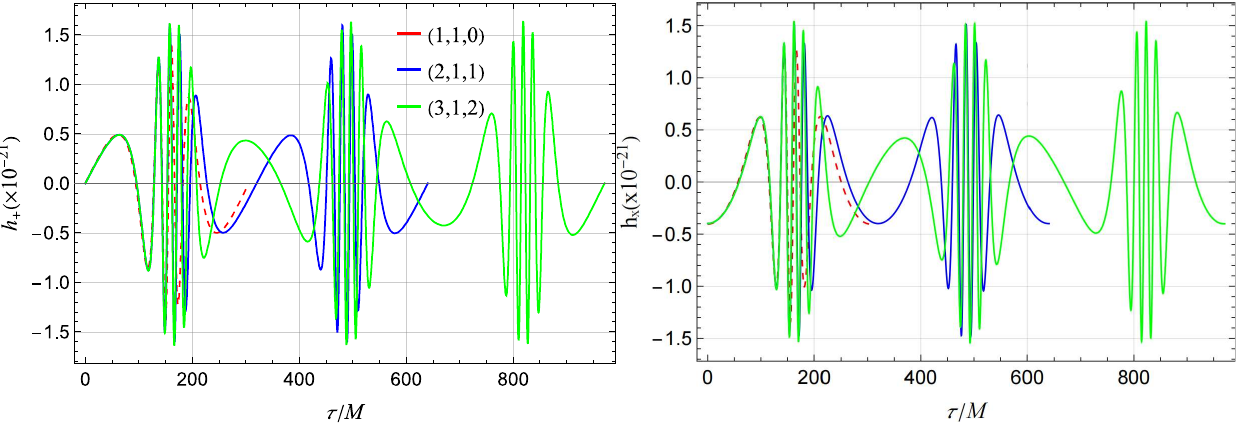}
		\caption{Different gravitational waveforms for $\xi=1$ and $Q=0.5$.}
		\label{fig:77}
	\end{figure*}
	Next, we investigate the effect of $Q$ and $\xi$ on GW radiation. Assume the mass of the test particle is $m = 10 M_{\odot}$, while the mass of the massive GW (i.e., the non-minimally coupled EYM BH) in the EMRI system is $M = 10^7 M_{\odot}$. The luminosity distance is taken as $D_L = 200$ Mpc, and the orbital inclination and orbital dimension angle are taken as $\iota = \frac{\pi}{4}$ and $\zeta = \frac{\pi}{4}$, respectively.
	
\begin{figure*}
		\centering
		\includegraphics[width=1\linewidth]{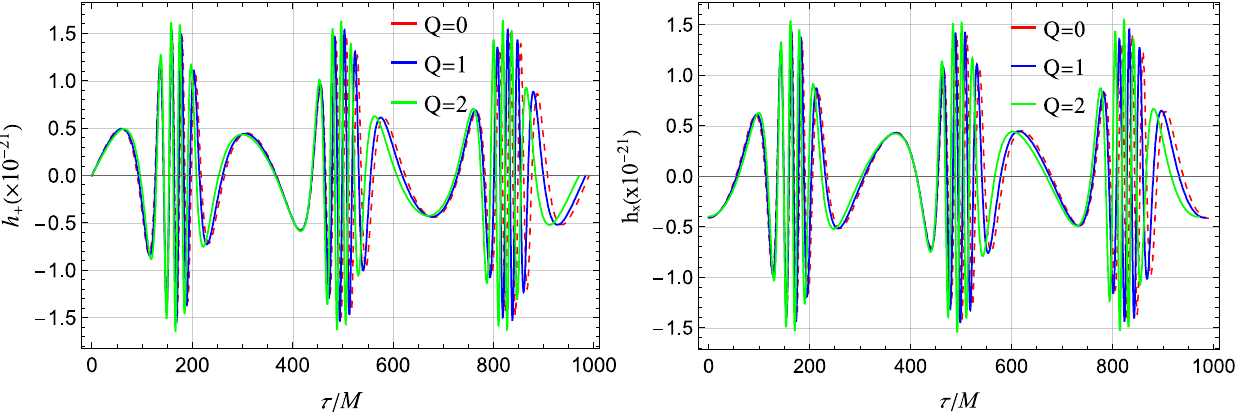}
		\caption{Gravitational waveforms for the periodic orbit $(3,1,2)$ with $\xi=1$ and different magnetic charge values $Q$.}
		\label{fig:78}
	\end{figure*}

\begin{figure*}
		\centering
		\includegraphics[width=1\linewidth]{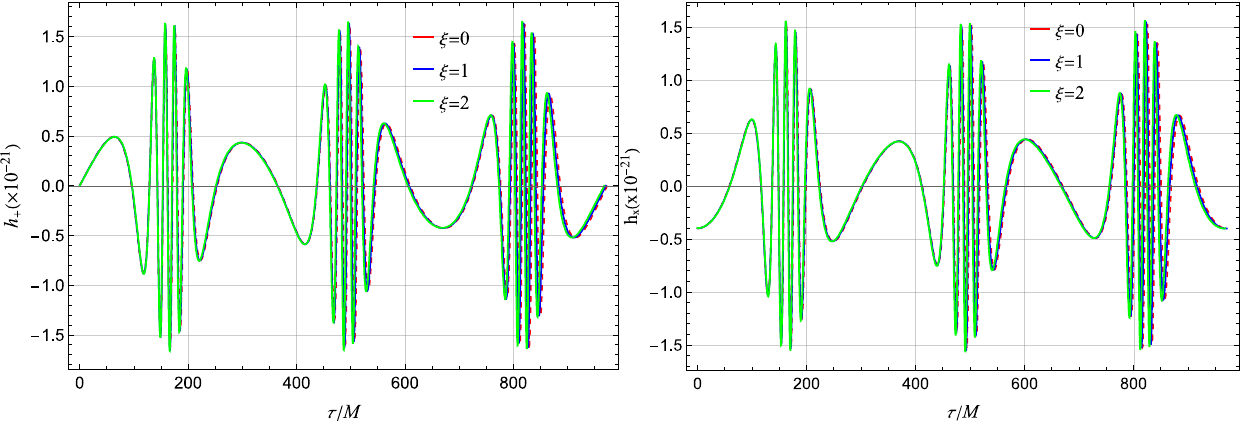}
		\caption{Gravitational waveforms for $(3,1,2)$ orbit with $Q=0.5$ and different $\xi$ values.}
		\label{fig:79}
	\end{figure*}

	As shown in Fig. \ref{fig:77}, Fig. \ref{fig:78}, and Fig. \ref{fig:79}, we plot the gravitational waveforms under different parameters. It can be clearly observed that the zoom-whirl behavior is a characteristic feature \cite{Vargas:2023gvd,Tu:2023xab,Lin:2022ysj}. A longer inspiral phase corresponds to the leaf-like structure of a circular orbit; at this time, the test particle is far from the BH, and the gravitational field is relatively weak. Conversely, rapid zooming corresponds to the whirl phase of a circular orbit; at this time, the particle is close to the BH, and the gravitational field is relatively strong.
	
	Fig. \ref{fig:78} investigates the influence of different $Q$ on the GWs. The red curve represents the Schwarzschild BH scenario. The comparison showsat the presence of ma thgnetic charge can cause a phase shift and simultaneously reduce the period and increase the amplitude. The greater the value of $Q$, the more prominent this effect becomes, which is consistent with Fig. \ref{fig:66}. In Fig. \ref{fig:79}, we fix $Q = 0.5$ and plot the GWs for different $(3, 1, 2)$ orbits. The RN BH case is depicted by the red dashed line. The results show that the presence of $\xi$ similarly causes a phase shift, an increase in amplitude, and a decrease in period. This result is consistent with the findings in Table \ref{tab:orbital_comparison}.
	In summary, the presence of $Q$ and $\xi$ will affect the gravitational waveforms. These differences serve as important criteria for distinguishing non-minimally coupled EYM BHs from RN and Schwarzschild BHs. Although we investigate the radiation GWs under the adiabatic approximation, the single-period EMRI system is sufficient to analyze the zoom-whirl behavior characteristics of the test particle's orbit. We can predict that in the future, when non-adiabatic effects are not ignored, the GW radiation model in the background of non-minimally coupled EYM BHs can be more accurately explored. Future space facilities such as Tianqin, Taiji, and LISA will provide high-precision observational data support to test the theoretical models of non-minimally coupled EYM BHs. Through precise measurements of the GW signals from EMRIs, it is hoped to directly detect the quantum structure of BH spacetime, thereby pushing strong field gravitational physics into the quantum era \cite{Liu:2020ijk,Burko:2020gse,Haiman:2017szj,Shabbir:2026qlh}.

	\section{\label{sec:6}Discussion and conclusions}
	In our study, we studied the GW radiation from periodic orbits in the background of a non-minimally coupled EYM BH. We investigated the influence of $Q$ and $\xi$ on the orbital characteristics (such as the zoom-whirl behavior) and the resulting GWs. The results show that the spacetime geometry corrections represented by $Q$ and $\xi$ lead to significant and observable differences in orbital frequency, stability, and GW radiation compared to the classical RN and Schwarzschild BHs. This provides a theoretical basis for using future gravitational wave data to test such ``hairy" BH models and constrain their theoretical parameters.
	
	Specifically, we treated the non-minimally coupled EYM BH as a supermassive BH and the compact star as a test particle. By solving the geodesic equations, we derived the effective potential as well as the criteria for bound orbits and periodic orbits. Among them, $Q$ is an important ``hairy parameter", whose value must satisfy the physical conditions for the presence of BHs, i.e., not exceeding the critical value determined by Fig. \ref{fig:13}, to avoid the appearance of naked singularities. The main content of this article is to select different values of $\xi$ and $Q$ (where $\xi \le \xi_c$) within the BH range determined by Fig. \ref{fig:13} for investigation.
	
	We explored the bound orbits of test particles in the spacetime of the non-minimally coupled EYM BH, mainly including two special characteristics: the bound orbits with the largest $E$ and the bound orbits with the largest $L$. We analyzed the results as follows: as $Q$ and $\xi$ increase, the effective potential well deepens, and the radius of the bound orbits decreases for both types of orbits. In contrast to the RN and Schwarzschild BHs, the test particle tends to move along orbits closer to the central BH. Bound orbits appear when the test particle travels between ISCO and MBO. Accordingly, we investigated the correlation between the orbital $L$ and $E$ in the $E-L$ plane. The results show that as $Q$ and $\xi$ increase, the $E-L$ space shifts to the left, but the magnetic effect is greater, and  has a ``suppression" effect on the magnetic field. The larger $\xi$ is, the more obvious the effect. As $\xi$ increases, the different $E-L$ spaces of the $Q$ will gradually converge towards the Schwarzschild BH situation.
	
	Furthermore, we used the rational number $q$ to explore the changes in the bound orbits within their allowed $E-L$ space: the rational number $q$ increases slowly with the increase of $E$ (decrease of $L$). The joint regulation of the magnetic charge $Q$ and $\xi$ makes $E$ and $L$ required for the bound orbits of the test particle decrease, as shown in Fig. \ref{fig:64} and Fig. \ref{fig:65}. For the bound orbits of the test particle controlled by the rational number $q$, the orbits for $Q=0$ are different from those for the Schwarzschild BH. At the same time, the bound orbits for the non-minimally coupled EYM black hole and the RN BH backgrounds are also different, as shown in Table \ref{tab:orbital_comparison}. This provides a distinguishing basis for differentiating non-minimally coupled EYM BHs, Schwarzschild BHs, and RN BHs.
	
	We explored the radiation gravitational waveforms in a single period by setting appropriate parameters for the extreme mass ratio inspiral system. We can clearly observe the zoom-whirl behavior characteristics. Investigating the different precession numbers $h_{+}$ and $h_{\times}$ under different magnetic charges and non-minimal coupling constants is to distinguish the simultaneous influence of the two parameters on the non-minimally coupled EYM black hole spacetime geometry. We found that $\xi$ and the $Q$ can both cause phase shifts, increase the amplitude, and decrease the period. This may provide a basis for supporting the later research on its internal structure and orbital evolution.
	
	In summary, our research results show that there is a non-trivial interaction and coordination relationship between $\xi$ and $Q$ effect: the $Q$ will cause the orbit to shrink and the stability condition to change, while $\xi$ not only enhances this trend but also exhibits a suppression effect on the magnetic field to a certain extent, making the orbital dynamics of different values of the non-minimally coupled EYM BH differ from the characteristics of the Schwarzschild BH. These characteristics provide a clear fingerprint information for the non-minimally coupled EYM BH to distinguish it from the Schwarzschild and RN BHs. Our work will provide a theoretical basis and specific observational guidelines for using EMRI GW signals to detect the ``hairy" structure of BHs. At the same time, in the future, we will take into account the significant differences in the self-force in the non-minimally coupled EYM BH spacetime, draw the orbital dynamics and GW radiation more accurately, and explore deeper physical laws.
	
	\section{acknowledgements}
	This work was supported by Guizhou Provincial Basic Research Program (Natural Science) (Grant No.QianKeHeJiChu[2024]Young166), the National Natural Science Foundation of China (Grant No.12365008), the Guizhou Provincial Basic Research Program (Natural Science) (Grant No.QianKeHeJiChu-ZK[2024] YiBan027 and QianKeHeJiChuMS[2025]680), the Guizhou Provincial Major Scientific and Technological Program XKBF (2025) 010 (Hosted by Professor Xu Ning), the Guizhou Provincial Major Science and Technological Program XKGF (2025) 009 (Hosted by Professor Xiang Guoyong) and Guizhou Provincial Major Scientific and technological Program (Teacher Fan Lu Lu moderated).
	
	\bibliography{ref}
	\bibliographystyle{apsrev4-1}

\end{document}